\documentclass[12pt]{iopart}
\RequirePackage{ifpdf}
\ifpdf
\usepackage[pdftex]{graphicx}
\else
\usepackage[dvips]{graphicx}
\fi
\usepackage{amssymb}
\usepackage{bm}
\usepackage{xspace}
\RequirePackage{url}

\newcommand{\Eqref}[1]{Eq.~\ref{#1}}

\newcommand{\Figref}[1]{Fig.~\ref{#1}}

\newcommand{\subgraph}{\in}
\newcommand{\GnuplotRotatePS}{\ifpdf 0.0 \else -90 \fi}

\newcommand{\Graph}   [1]{\mathcal{#1}}
\newcommand{\Conn}    [1]{C(\Graph{#1})}
\newcommand{\Edge}    [1]{E(\Graph{#1})}

\renewcommand{\Im}[1]{\mathrm{Im}(#1)}

\newcommand{\Edges}   [1][]{\mathsf{E_{\Graph{#1}}}}
\newcommand{\Clusters}[1][]{\mathsf{C_{\Graph{#1}}}}
\newcommand{\Vertices}[1][]{\mathsf{V_{\Graph{#1}}}}

\newcommand{\Vertex}[1]{V(\Graph{#1})}
\newcommand{\Rank}  [1]{r(\Graph{#1})}
\newcommand{\GDos}[1]{\Gamma_{\Graph{#1}}(\Clusters , \Edges)}
\newcommand{\MaxE}{2N}
\newcommand{\MaxV}{N}

\begin{document}

\title{The number of link and cluster states: the core of the 2D $q$
  state Potts model} 

\author{J. Hove}

\address{%
  Institute of physics and technology \\
  5020 Bergen\\}
\date{\today}
\ead{hove@ift.uib.no}


\begin{abstract}
  Due to Fortuin and Kastelyin the $q$ state Potts model has a
  representation as a sum over random graphs, generalizing the Potts
  model to arbitrary $q$ is based on this representation. A key
  element of the Random Cluster representation is the combinatorial
  factor $\Gamma_{\Graph{G}}(\Clusters,\Edges)$, which is the number
  of ways to form $\Clusters$ distinct clusters, consisting of totally
  $\Edges$ edges. We have devised a method to calculate
  $\Gamma_{\Graph{G}}(\Clusters,\Edges)$ from Monte Carlo
  simulations. 
\end{abstract}
\maketitle

\section{Introduction}
The Potts model\cite{Wu:1982} is one of the most studied models in
statistical physics. The traditional representation of the model is in
terms of the Hamiltonian
\begin{equation}
  \label{PottsH}
  H = -J\sum_{\langle i,j\rangle} \delta(\sigma_i,\sigma_j),  
\end{equation}
where the spins $\sigma_i$ are integer values $\sigma_i \in [1\ldots
q]$, the sum $\langle i,j \rangle$ is over nearest neighbours. The $q$
is a parameter of the model. The model is typically defined on a
regular lattice in $d$ dimensions, but can in general be defined on
any graph.

For $d \ge 2$ the model sustains a order-disorder transition , in
$d=2$ the critical coupling is $\beta_c = \ln(1 + \sqrt{q})$. For
$\beta > \beta_c$ the $q$-fold permutation symmetry of \Eqref{PottsH}
is broken, and one of the $q$ different groundstates has been singled
out.  For $q=2$ the model is the familiar Ising model, which has a
second order transition, but with increasing $q$ the excited states
have relatively more entropy and for $q > q_c$ the transition is first
order. For $d=2$ the phase transition changes order at $q_c = 4$
\cite{Baxter:1973,Nienhuis:1979}, for $d=3$ the exact value is not
known, but the most recent estimate based on Monte Carlo simulations
is $q_c \approx 2.35$\cite{Hartmann:2005}.

The Hamiltonian \Eqref{PottsH} is only defined for integer $q$,
however due to an elegant transformation by Fortuin and Kastelyn (KF)
the partition function of the $q$ state Potts model can be written as
a correlated percolation problem, the socalled Random Cluster (RC)
model\cite{Fortuin:1972}. In the RC representation $q$ enters as an
ordinary variable, and can attain any scalar value. Apart from
extrapolation/interpolation from integer $q$ results, all (numerical)
studies of the noninteger $q$ properties of the Potts model are based
on the RC representation, this also applies to the current paper.
Properties of the Potts model with noninteger $q$ have been
extensively studied using transfer matrix\cite{Blote:1982} techniques.
Recently also MC simulations have been used. The latter come in two
categories; either a technique is based on the RC measure to simulate
directly at an arbitrary
$q$\cite{Hartmann:2005,Gliozzi:2002,Deng:2004}, or alternatively the
results are reweighted to arbitrary $q$ after the simulation is
complete\cite{Gliozzi:2002,Weigel:2002}.

The rest of this paper is organized as follows: In section
\ref{GraphT} we introduce some key elements of graph theory, and how
concepts from graph theory can be applied in statistical physics; in
particular to the Potts model. In section \ref{Algorithm} we introduce
and describe an algorithm which can be used to ``reweight'' Potts
model simulations to arbitrary $q$. Section \ref{Results} is devoted
to results, both to show the correctness of the approach and also to
study real $q$ properties which are not easily studied by ordinary MC
simulations.

\section{Graph theory and the Potts model}
\label{GraphT}
\begin{figure}[htbp]
\centerline{\scalebox{0.50}{\rotatebox{0.00}{\includegraphics{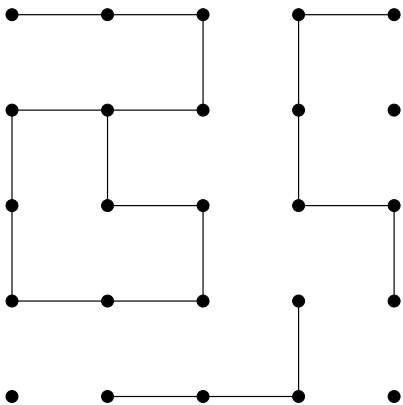}}}}
\caption{\label{Graph} The sites of $5 \times 5$ lattice, and links
  connecting some of the sites. Together these sites and links
  constitute a \emph{graph}. This particular graph has $\Vertices[G] =
  25$, $\Edges[G] = 20$, six connected components ($\Clusters[G] = 6$) and a rank 
  $\Rank{G} = \Vertices[G] - \Clusters[G] = 19$.}
\end{figure}

An (undirected) graph $\Graph{G}$ is a collection of vertices
$\Vertex{G}$, along with a set of edges $\Edge{G}$ connecting the
vertices\cite{Wilson:1985:book}. A subgraph $\Graph{G'} \subgraph
\Graph{G}$ is a collection of vertices and edges such that
$\Vertex{G'} \subgraph \Vertex{G}$ and $\Edge{G'} \subgraph \Edge{G}$.
The rank of a graph is denoted by $\Rank{G}$ and given by
\begin{equation}
  \label{RankDef}
  \Rank{G} = \left| \Vertex{G} \right| - \Conn{G},
\end{equation}
where $\left| \Vertex{G} \right|$ is the number of vertices and
$\Conn{G}$ is the number of connected components.
Observe that also isolated single vertices constitute connected
components when evaluating the rank of a graph. \Figref{Graph} shows a
simple graph and illustrates the necessary concepts. From now on we
will use the symbols $\Edges[G]$, $\Clusters[G]$ and $\Vertices[G]$ to
denote the number of edges, clusters and vertices in a graph
$\Graph{G}$, when there is no ambiguity we will omit the index $\Graph{G}$.

By assigning scalar properties to sites and bonds one can define
different \emph{graph polynomials}. One of the most general graph
polynomials is the Tutte or Di-Chromatic polynomal
$T_{\Graph{G}}(x,y)$\cite{Tutte:1954,Brylawski:1992:book}:
\begin{equation}
  \label{TutteDef}
  T_{\Graph{G}}(x,y) = \sum_{E \in \Edge{G}} \left(x - 1\right)^{r(E) - \Rank{\Graph{G}}} \left(y - 1 \right)^{\Edges - \Rank{G}}.
\end{equation}
The sum in \Eqref{TutteDef} is over all edge configurations of the
graph $\Graph{G}$ (i.e. \emph{spanning} subgraphs). Here $x$ is a
scalar property assigned to the vertex set, and $y$ a property
assigned to the edges; as indicated in \Eqref{TutteDef} we will only
consider the situation of spatially constant $y$, but the general
definition of the Tutte polynomial allows for a set $\left\{ y
\right\}$ of edge properties.  Many other polynomials can be found as
suitably rescaled evaluations of the Tutte
polynomial\cite{Welsh:1993:book}:
\begin{eqnarray}
  \label{Rp}  R_{\Graph{G}}(p) &=& (1-p)^{\Edges - \Vertices + 1}p^{\Vertices - 1} T_{\Graph{G}}\left(1, \frac{1}{1-p}\right) \\
  \label{Pq}  P_{\Graph{G}}(q) &=& (-1)^{r(E)}q^{\Clusters}T_{\Graph{G}}(1-q,0)\\
  \label{Zq}  Z_{\Graph{G}}(q,v) &=& q v^{\Vertices -  1}(v+1)^{-\Edges} T_{\Graph{G}} \left(\frac{q + v}{v}, v + 1
  \right), \quad v = \frac{p}{1 - p} = e^{\beta J} - 1.
\end{eqnarray}
$R_{\Graph{G}}(p)$ is the \emph{reliability} polynomial, closely
related to the (bond) percolation problem. $P_{\Graph{G}}(q)$ is the
\emph{chromatic} polynomial, and denotes the number of ways the
vertices in $\Graph{G}$ can be colorized with $q$ different colors, so
that no adjacent vertices share the same color. The chromatic
polynomial coincides with the $T \to 0$ limit of the partition
function of the \emph{anti}ferromagnetic Potts model. Finally $Z(q,v)$
is the partition function of the $q$ state Potts model. Observe the
quantity $v$ in \Eqref{Zq}, in this context this is the most
convenient temperature variable.

The FK transformation is the key to identify $Z(q,v)$ with the Tutte
Polynomial\cite{Fortuin:1972}. The actual transformtion is in terms of
the complete partition function, hence it is not possible to identify
a spin state with a corresponding RC state uniquely, see however Ref.
\cite{Sokal:1988} for an exposition in terms of a mixed bond-spin
model which elucidates the connection. $Z_{\mathrm{RC}}(p,q)$ is a
function of two variables: a probability $p$ to occupy an edge, and a
$q$, where $\ln q$ resembles a cluster entropy. The RC partition
function is built up as follows: (1) each configuration
$E'(\Graph{G})$ of edges gets a ``Boltzmann''-weight
$p^{\Edges'}(1-p)^{\Edges - \Edges'}$, (2) the weight is multiplied by
an entropic factor $q^{\Clusters'}$, (3) all configurations $E'(\Graph{G})$ are summed
over. This finally gives the RC partition function
\begin{equation}
  \label{ZFK}
  \fl \qquad
  Z_{\mathrm{RC}}(q,p) = \sum_{E'(\Graph{G}) \subgraph E(\Graph{G})} p^{\Edges'} (1-p)^{\Edges - \Edges'} q^{\Clusters} = 
  \sum_{\Clusters = 1}^{\Vertices} \underbrace{\sum_{\Edges'=0}^{\Edges} \Gamma_{\Graph{G}}(\Clusters,\Edges)  p^{\Edges} (1-p)^{\Edges - \Edges'}}_{a_C(p)} q^{\Clusters}, 
\end{equation}
The $p$ in \Eqref{ZFK} is the probability to occupy an edge, for the
RC model this is an arbitrary number, however to make contact with the
$q$-state Potts model at coupling $\beta$, we must have $p = 1 -
e^{-\beta J}$. As indicated in \Eqref{ZFK} the partition function can
bee seen as a polynomial in $q$, with $p$ dependant coeffiecients. In
section \ref{complex:zero} we will use this to determine the zeroes of
the partition function in the complex $q$ plane.

Using the combinatorial factor $\Gamma_{\Graph{G}}(\Clusters,\Edges)$
to express the sum is the key element in \Eqref{ZFK}. This factor is
simply the number of ways to form $\Clusters$ connected components
with $\Edges$, on the underlying graph $\Graph{G}$. This is a purely
combinatorial/geometric property which can in principle be calculated
without any reference to a particular model of statistical physics. On
the other hand all physical properties are contained in $\GDos{G}$.
\Eqref{ZFK} also highlights that the Potts model has a common
structure independent of $q$, even though the physical properties vary
significantly with $q$. In addition to facilitating the study of the
Potts model for arbitrary $q$, the FK representation also serves as
the theoretical underpinning of the Swendsen-Wang algorithm for spin
models\cite{Sokal:1988,Swendsen:1987}.

An important topic in computer science is a formal demarcation of
tractable and intractable problems. The socalled $\# P$ complete
problems are counting problems which are essentially intractable.
Obtaining the partition function of (discrete) system belong to this
category\cite{Hartmann:2005,Welsh:1990}. Due to this intractability
good approximative techniques is essential; the Monte Carlo technique
is one such approach. Also in computer science the use of Monte Carlo
techniques to approach $NP$ and $\#P$ complete problems, has been
popular, see eg. \cite{Jerrum:1996}. Computer scientists Jerrum and
Sinclair have devised efficient Monte Carlo algorithms (FPRAS) to
determine the partition functions of both 2D monomer-dimer system, and
the 2D Ising model\cite{Jerrum:1989,Jerrum:1993}. Hence the study of
the RC and related problems is of interest to scientist from widely
different fields.

\section{Algorithm}
\label{Algorithm}
The probability $P(\epsilon)$ to find a system in a state with energy
$\epsilon$ is proportional to $g(\epsilon)e^{-\beta \epsilon}$, where
$g(\epsilon)$ is the density of states at energy $\epsilon$. That
$P(\epsilon)$ can be written in this manner is the foundation of
ordinary $\epsilon - \beta$ reweigting\cite{Ferrenberg:1989}. In the
formulation \Eqref{ZFK} $(p,\Edges)$ and $(q,\Clusters)$ are
``conjugate'' variable pairs; alas $\GDos{G}$ can be used to reweight
to arbitrary $q$ and $p$; from now on we will mostly use $\beta$ in the
text, but it should be understood that the relation $p = 1 - e^
{-\beta J}$ applies throughout. In the remainder of this section we
will present an algorithm to estimate $\GDos{G}$ from simulations at
different $p$ and $q$. An algorithm based on the same principle was
presented by Weigel et. al.  in Ref. \cite{Weigel:2002}, and just
recently Hartmann has presented an algorithm based on only $(q,\Clusters)$
reweighting\cite{Hartmann:2005}.

The algorithm presented here is general, and will apply to any graph.
However for ease of notation we have specialized to a two dimensional
square lattice with a total of $N = L\times L$ sites, and $2N$ edges.
The Gibbs probability to find \emph{any} state with $\Clusters$ components and
$\Edges$ edges is given by\cite{Welsh:1993:book}:

\begin{equation}
  \label{PCompEdge}
  P_{\Graph{G}}(\Clusters,\Edges) = \frac{\GDos{G} p^{\Edges} q^{\MaxE - \Edges} q^{\Clusters}}{Z_{\Graph{G}}(q,\beta)}.    
\end{equation}

To estimate $\GDos{G}$ we need to generate states distributed
according to \Eqref{PCompEdge}. We have done this by using the
Swendsen-Wang\cite{Swendsen:1987} algorithm on the $q$ state Potts
model, with integer $q$. However, one could equally well have used an
algorithm generating RC states directly\cite{Gliozzi:2002,Deng:2004},
or alternatively a combination. During the simulation at $\mu =
(q,\beta)$ a histogram $h_{\mu}({\Clusters},\Edges)$ is collected.
From the histogram $h_{\mu_0}({\Clusters},\Edges)$ we can in principle
estimate $\GDos{G}$ from \Eqref{PCompEdge}
\begin{equation}
  \label{GammaEstI}
  \hat{\Gamma}_{\mu_0}({\Clusters},\Edges) = e^{\xi_{\mu_0}} h_{\mu_0}({\Clusters},\Edges) p_0^{-\Edges}q_0^{-(\MaxE-\Edges)} q_0^{-{\Clusters}}, 
\end{equation}

where $\xi_{\mu_0}$ is an (undetermined) normalization constant.
$\GDos{G}$ is independent of $\mu$, however the estimator in
\Eqref{GammaEstI} has been given index $\mu_0$ to indicate that it is
based on results sampled at these couplings. The estimator
\Eqref{GammaEstI} is formally correct, but only applicable in a narrow
range around the mean values $\langle {\Clusters} \rangle_{\mu_0}$ and $\langle
\Edges \rangle_{\mu_0}$. By combining results obtained at different
$\beta$ and $q$ we can get an estimate for $\GDos{G}$ which is valid
for a wide range of ${\Clusters}$ and $\Edges$ values. A series of $N$
histograms obtained at couplings $\mu_1,\mu_2,\ldots,\mu_N$ can be combined as

\begin{equation}
  \label{GammaEstII}
  \hat{\Gamma}_{\Graph{G}}({\Clusters},\Edges) = \sum_i^{N} w_i({\Clusters},\Edges) \cdot \hat{\Gamma}_{\mu_i}({\Clusters},\Edges),
\end{equation}
where the weight factor $w_i({\Clusters},\Edges)$ is given by 
\begin{equation}
  \label{weight}
  w_i({\Clusters},\Edges) = \frac{h_{\mu_i}({\Clusters},\Edges)}{\sum_k h_{\mu_k}({\Clusters},\Edges)}.
\end{equation}

The normalization constants $\xi_i, i > 1$ are determined by
maximizing, the (weighted) overlap between (the logarithm of) the
estimates $\hat{\Gamma}_{\mu_i}({\Clusters},\Edges)$. Mathematically
this amounts to minimizing
\begin{eqnarray}
  \label{Chi}
  \fl
  \chi^2 = \sum_i \sum_{j > i} \sum_{{\Clusters},\Edges} 
  h_{\mu_i}({\Clusters},\Edges)h_{\mu_j}({\Clusters},\Edges) \times \nonumber \\
  \bigg( 
  \underbrace{\left(\xi_i + \ln h_{\mu_i}({\Clusters},\Edges) - {\Clusters}\ln q_i\right) - 
              \left(\xi_j + \ln h_{\mu_j}({\Clusters},\Edges) -  \Clusters \ln q_j\right)}_{\ln \hat{\Gamma}_{\mu_i} - \ln \hat{\Gamma}_{\mu_j}} 
  \bigg)^2,
\end{eqnarray}
with $\xi_1$ initially fixed at an arbitrary value.  The final
normalization constant $\xi_1$ is determined by the overall
normalization
\begin{equation}
  \label{GammaNorm}
  \sum_{{\Clusters},\Edges'} \Gamma_{\Graph{G}}({\Clusters},\Edges') = 2^{\Edges}.  
\end{equation}
The actual solution of the minimization problem \Eqref{Chi} is found
as the solution of a system of linear equations. As long as all the
histograms $h_{\mu_i}(\Clusters,\Edges)$ have finite overlap with at
least one other histogram $h_{\mu_j}(\Clusters,\Edges)$ the solution
will be found. The method is a generalization of an existing algorithm
to determine the density of states $g(\epsilon)$\cite{Alves:1990,Hove:2004}.

Due to the nonlinear nature of the algorithm it is difficult to
calculate errors by the use of error-propagation.  Furthermore the
estimation of $\Gamma_{\Graph{G}}({\Clusters},\Edges)$ is quite time
consuming, hence computer-intensive methods like Jack-Knife and
Bootstrap are not very suitable. In the current paper error estimates
have been calculated by comparing the results from independent
simulations.

\section{Results}
\label{Results}
\subsection{Basic thermodynamic results}
\begin{figure}[htbp]
  \centerline{\scalebox{0.70}{\rotatebox{\GnuplotRotatePS}{\includegraphics{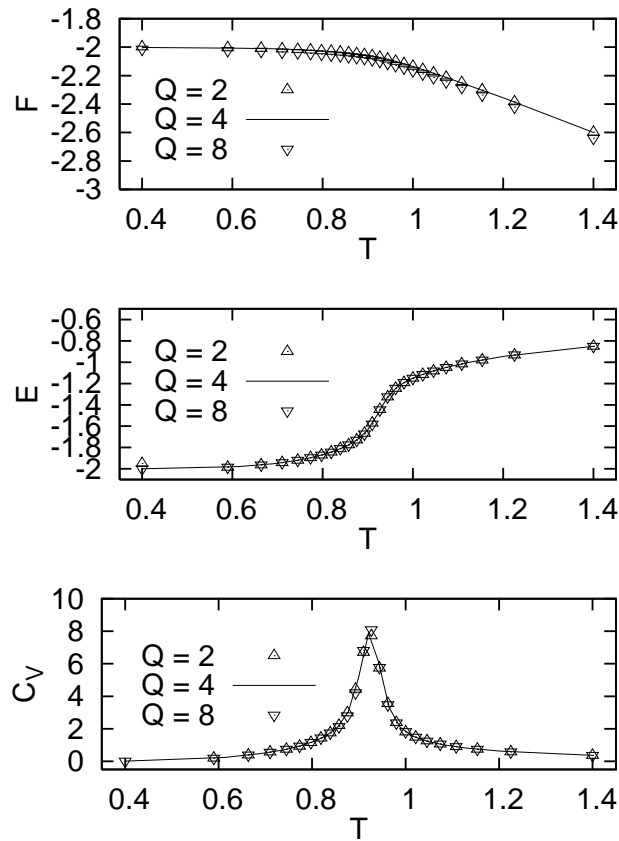}}}}
  \caption{\label{Fig:ThermoQ4}This figure shows from top to bottom free energy, internal energy and spesific heat 
    for the $q=4$ Potts model, system size is $16 \times 16$. The solid line shows result obtained from a
    simulation at $q = 4$, the symbols show results ``reweighted'' from $q=2$ and $q=8$ respectively.}
\end{figure}

In this section we will show how simulations performed at one value
$q_1$ can be reweighted to another $q_2 \neq q_1$.
\Figref{Fig:ThermoQ4} shows thermodynamics for a $q = 4$ Potts model.
The solid line is data obtained at $q = 4$, and the symbols represent
results reweighted from $q=2$ and $q=8$ respectively.

\subsection{The average trajectory in clusters - links space}
\begin{figure}[htbp]
  \centerline{\scalebox{0.55}{\rotatebox{0.00}{\includegraphics{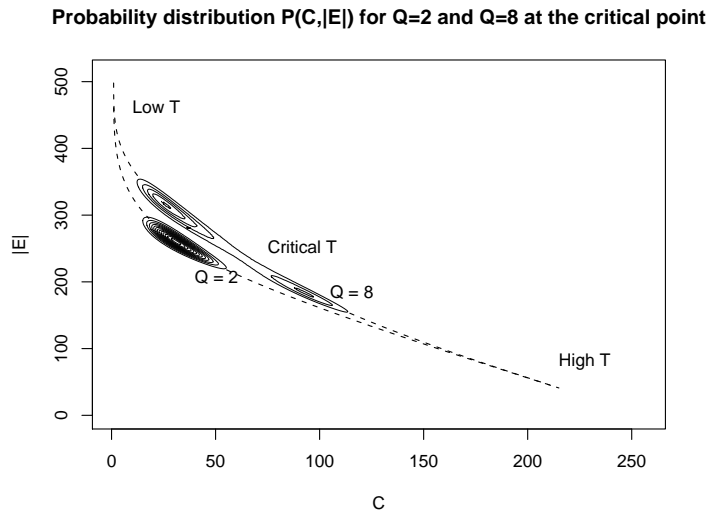}}}}
  \caption{\label{CLDist}Contour plot of the density $P({\Clusters},\Edges)$ at the
    critical point, for $q=2$ and $q=8$ for a $16 \times 16$ lattice. The dashed lines 
    show $\langle {\Clusters} | \Edges \rangle$, which corresponds to the path
    followed in ${\Clusters},\Edges$ space when temperature is
    varied.}
\end{figure}

In the Random Cluster formalism the state of the system is given by
${\Clusters}$ and $\Edges$, and it is interesting to see how these
quantities evolve when the Potts model parameters $\beta$ and $q$ are
varied. For a fixed value of $\Edges$ the \emph{conditional}
probability $P({\Clusters}|\Edges)$ is \emph{independent} of
$\beta$; hence we can easily plot the mean path the system will
follow in $({\Clusters},\Edges)$ space. In \Figref{CLDist} we show the
conditional mean
\begin{equation}
  \label{MeanC}
  \langle {\Clusters} | \Edges \rangle = \frac{\sum_{{\Clusters}}{\Clusters} \cdot \Gamma({\Clusters},\Edges)q^{\Clusters}}{\sum_{{\Clusters}} \Gamma({\Clusters},\Edges) q^{\Clusters}},
\end{equation}
along with the contours of $P(\Clusters,\Edges)$ at the critical
coupling, for two different values of $q$.  As we can see from
\Figref{CLDist} the $q$ behaviour of ${\Clusters}$ and $\Edges$ can
conveniently be divided in three regions: (1) a low $T$ region where
$\langle {\Clusters}|\Edges\rangle \approx 1$ quite independent of
$\Edges$, a high $T$ region where $\langle {\Clusters} |\Edges \rangle
\gtrsim \MaxV - \Edges$ and an intermediate region containg the
critical point. It is only in the intermediate region there is
significant $q$ dependence.

The contours in \Figref{CLDist} show the probability density
$P({\Clusters},\Edges)$ at the critical point, for $q=2$ and $q=8$.
The $q$ ``reweighting'' has similar limitations as ordinary thermal
reweighting, the statistics is best at the original $q$ value, and can
not be extended to regions of $({\Clusters},\Edges)$ space which have
not been sampled. As we can from \Figref{CLDist} the overlap between
the $q=2$ and $q=8$ results is very small; hence reweighting between
these two $q$ values would give unreliable results. 

\begin{figure}[htbp]
  \centerline{\scalebox{0.60}{\rotatebox{\GnuplotRotatePS}{\includegraphics{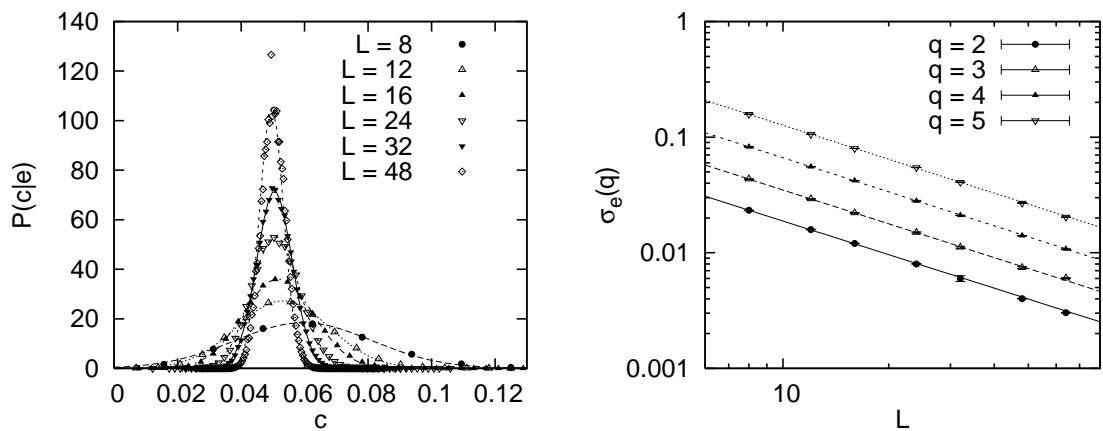}}}}
  \caption{\label{CFluct} The left figure shows the conditinal
    distribution $P(c|e)$ at $e = (1 - 1/(1 + \sqrt{q}))/L^2$,
    i.e. the critical link density, for the $q=3$ model.
  The right figure shows the width of the distribution $P(c|e)$ as a
  function of $L$, all the curves show a $L^{-1}$ decay. The curves for $q
    \ge 3$ have been shifted for clarity. }
\end{figure}

From \Figref{CLDist} we see that the fluctuations are quite
assymetric; they are much larger along the direction given by the mean
path \Eqref{MeanC} than orthogonal to it. The conditional distribution
function 
\begin{equation}
  \label{Eq:PcondC} P(\Clusters|\Edges) = \frac{\Gamma(\Clusters,\Edges)q^{\Clusters}}{\sum_{\Clusters}\Gamma(\Clusters,\Edges)q^{\Clusters}}
\end{equation}
is well described by a Gaussian with width $\sigma_{\Edges}(q)$. The
width scales with the number of sites as $N^{1/2}$, hence the
\emph{relative} fluctuations in the number of clusters scales as
$N^{-1/2}$ and consequently the system will follow an increasingly
well defined \emph{line} in $(\Clusters,\Edges)$ space when the system
size increases. \Figref{CFluct} shows the distribution of the cluster
density $\mathsf{c} = \Clusters/N$ for a given link density
$\mathsf{e} = \Edges/N$, and finite size scaling of the width of this
distribution, $\sigma_{\mathsf{e}}(q) = \sigma_{\Edges}(q)/N$.

\begin{figure}[htbp]
\centerline{\scalebox{0.50}{\rotatebox{\GnuplotRotatePS}{\includegraphics{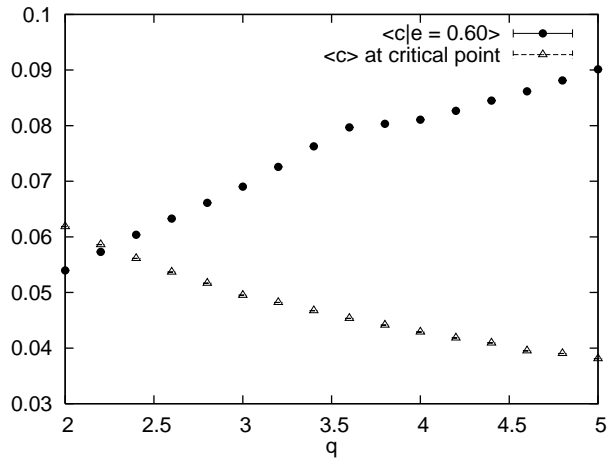}}}}
\caption{\label{meanC}The mean number density of clusters as a
  function of $q$, for a fixed density of links and at the ($q$
  dependant) \emph{critical} link density. The results in the figure are from a 
  $16 \times 16$ lattice.}
\end{figure}

In the RC model each cluster can be in $q$ different configurations,
hence we get an additive entropy contribution of $\ln q$ from every
cluster. Consequently we see that for a fixed number of links the
average number of clusters will increase with $q$. On the other hand
larger amount of entropy per cluster, means that for high $q$ entropy
will dominate the competetion between internal energy and entropy at a
lower number of clusters, and consequently \emph{at} the critical
point $\langle \Clusters \rangle$ decreases with increasing $q$. These
points are illustrated in \Figref{meanC}.

\subsection{Evaluation of the Tutte polonymial}
The Tutte polynomial can be defined in terms of a recursive
definition\cite{Welsh:1993:book}; which immediately leads to a simple
and exact algorithm for computation of $T_{\Graph{G}}(x,y)$.  However
this algorithm has exponential complexity, and is clearly not feasible
for anything but very small graphs.  Due to it's importance in many
different areas of mathematics and computer science, this has lead to
a large effort to find efficient approximate algorithms for evaluation
of the Tutte polynomial\cite{Alon:1995}.

\begin{figure}[htbp]
  \centerline{\scalebox{0.70}{\rotatebox{\GnuplotRotatePS}{\includegraphics{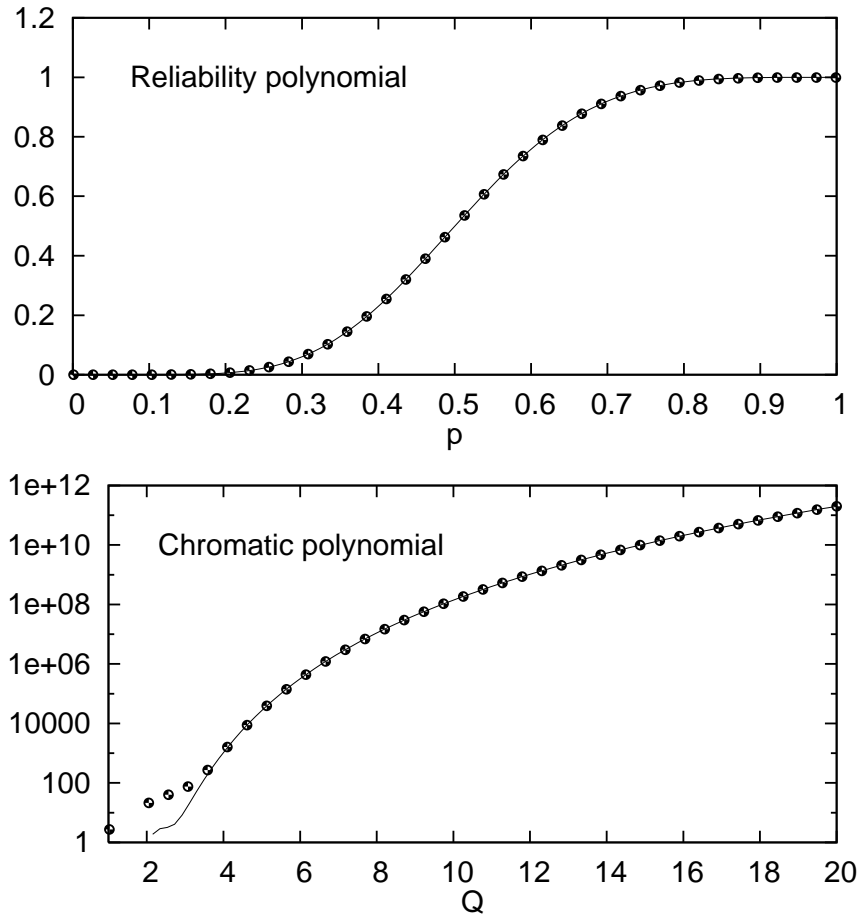}}}}
  \caption{\label{TutteFig}The reliability polynomial \Eqref{Rp} and chromatic polynomial \Eqref{Pq} for 
    a $3 \times 3$ lattice. The solid lines are exact results from the
    computer algebra system Maple, and the points come from our
    simulations. The very small system size considered is to limit the run-time of Maple}
\end{figure}

Using the algorithm presented here we can also estimate Tutte
polynomials, in \Figref{TutteFig} we show the reliability polynomial
and the Chromatic polynomial. With the current approach the running
time to determine the Tutte polynomial is governed by the running time
of the MC algorithm, and at least for $q \le 4$ the Swendsen-Wang
algorithm is rapidly mixing\cite{Gore:1999}.

When the arguments $x,y$ of the Tutte polynomial move a long way away
from the values used when sampling, the results become unreliable;
consult \Eqref{Zq} to see how $x$ and $y$ are related to the
parameters $q$ and $\beta$ of the Potts model. In particular for $x <
1$ and/or $y < 1$ the evaluation of $T(x,y)$ is difficult, because in
these regions the polynomial terms are oscillating and inaccurate
coefficients lead to large relative errors.

\subsection{Zeros in the complex $q$ plane}
\label{complex:zero}
The formulation of the partition function as a polynomial in $q$
allows for quite easy evaluation of the zeros of the partition
function in the complex $q$ plane. Properties of the complex $q$
zeroes have been investigated both analytically, and
numerically\cite{Sokal:2001}. According to the Yang-Lee view of
critical phenomena the critical point is characterized by zeros in the
complex $\beta$ plane pinching the real axis. The phase transition in
the Random Cluster model can be driven by both $\beta$ and $q$, we
should therefor see the same pinching of the real $q$ axis.

\begin{figure}[htbp]
  \centerline{\scalebox{0.65}{\rotatebox{\GnuplotRotatePS}{\includegraphics{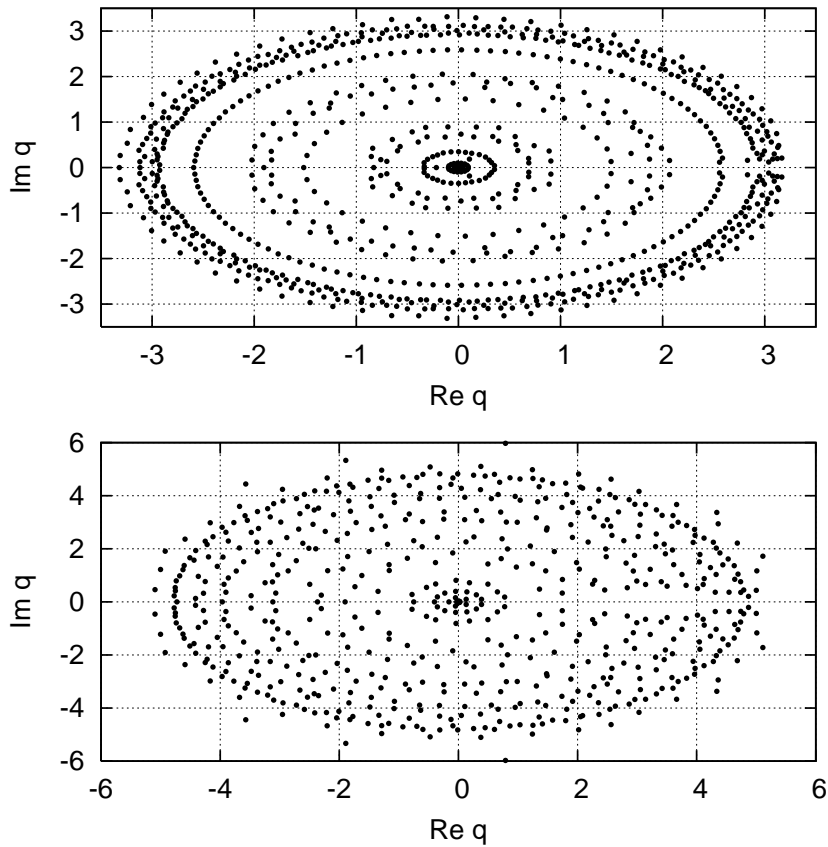}}}}
  \caption{\label{Fig:qRoots}The roots in the complex $q$ plane of the
    partition function $Z(v,q)$ at couplings $v = \sqrt{3}$ (top) and
    $v = \sqrt{5}$ (bottom). We observe
    that the zeroes close in on the critical $q$ values of 3 and 5.}
\end{figure}

The critical coupling is given by $\beta_c J = \ln(1 + \sqrt{q})$,
alternatively we find that for a \emph{fixed} $\beta$ the critical $q$
is given by
\begin{equation}
  \label{Eq:qc}
  q_c = \big(e^{\beta J} - 1\big)^2 = v^2.
\end{equation}
For the current discussion the temperature variable $v$, first
introduced in \Eqref{Zq} will be the most convenient.  Plotting the
zeros of $Z(v,q)$ we expect the zeros to pinch the real $q$ axis close
to the $q_c$ given by \Eqref{Eq:qc}, \Figref{Fig:qRoots} shows the
distribution of zeros in the complex $q$ plane for two different couplings.

\begin{figure}[htbp]
  \centerline{\scalebox{0.70}{\rotatebox{\GnuplotRotatePS}{\includegraphics{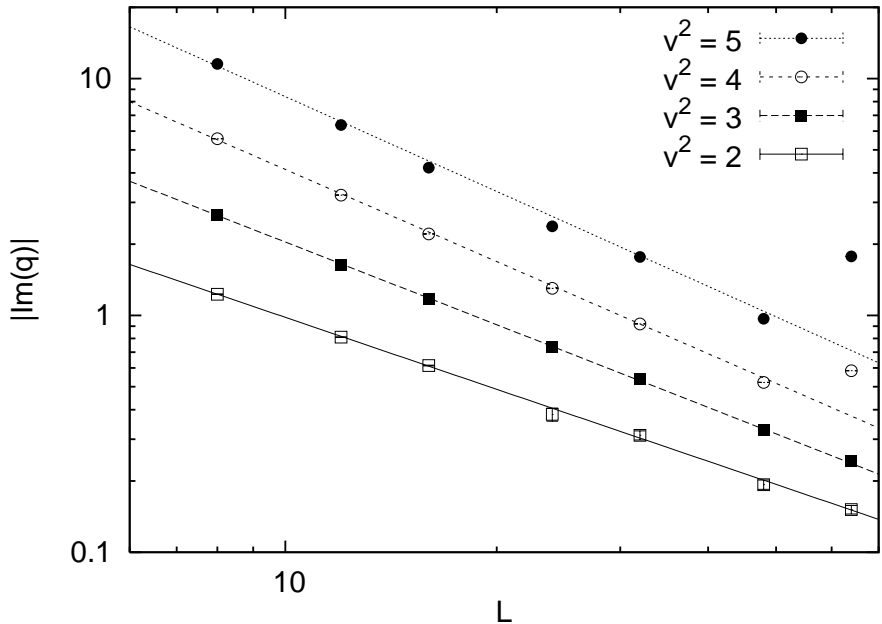}}}}
  \caption{\label{FSS:QZero}The plots show $|\Im{q}|$ for the zero
    closest to the real axis, as a function of system size $L$. The value of
    $v^2$ coincides with $q_c$. The error bars are generally smaller than the symbol size. The solid
    lines are least squares fits with slope, from top to bottom,
    $-1.3(1),-1.29(5),-1.159(9),-1.008(7).$ }
\end{figure}

If we denote the zero closest to $q_c$ with $q_c(L)$, we find that
$q_c(L)$ converges towards $q_c$ with increasing system size. To
determine which zero is indeed the ``critical'' one we have measured
distance $d(q_i,q_c)$ using both the ordinary metric $d_2(x,y) = |x -
y|$ and also $d_1(x,y) = |\mathrm{Im}(x) - \mathrm{Im}(y)|$. For
$
v^2 \lesssim 3.0$ the two methods select the same zero, whereas for
$v^2 \gtrsim 3.0$ different zeros are selected, and the real part of
the zero selected by $d_2$ jumps about randomly.  \Figref{FSS:QZero}
shows finite size scaling plots of the $|\Im{q}|$ (as determined by
using $d_1$) for the zero closest to the real $q$ axis. This should
scale as
\begin{equation}
  \label{ImQ:FSS}
  |\Im{q}| \sim L^{-\frac{1}{\nu}}.   
\end{equation}
For $q=2$ and $q=3$ this gives $\nu \approx 0.992(7)$ and $\nu \approx
0.863(7)$ which agree reasonably well with the exact values of $1$ and
$5/6 \approx 0.8333\ldots$. For $q=4$ we get $\nu \approx 0.77(3)$,
this is well above the exact value of $2/3$ + logarithmic
corrections. If we assume an \emph{effective exponent} for the first
order transition at $q=5$ we would expect $\nu=1/2$, whereas the
estimated value is $\nu = 0.77(6)$.

The reason that the quality of the $\nu$ estimates detoriate with
increasing $q$ is probably that the slope of the curve $\beta_c(q)$ is
reduced with increasing $q$. When the transition is driven by $q$ the
critical point is approached more and more tangentially. It seems
reasonable that this makes a precise determination of the critical
properties progressively more difficult. Furthermore the model has
limiting behaviour at $q=4$, with strong corrections to scaling;
consequently critical properties are notoriously difficult to
determine numerically at $q=4$\cite{Salas:1997}.

The zeroes are found using the MPSolve\cite{Bini:2001} package. To
determine the roots of $Z(v,q)$ in the complex $q$ plane is an
ill-posed problem. Firstly the coefficeints $a_C(p)$ (see \Eqref{ZFK})
vary over a \emph{wide} range, secondly finite sampling statistics
adds to the problem. In particular the states with $\Clusters \to N$
are typically not sampled at all. For independent simulations the
pattern of zeroes differs significantly from case to case, however the
location of the zero $q_c(L)$ shows much less fluctuations. The
results in \Figref{FSS:QZero} are the total of ten independent
simulations, and as we see the error bars are very small.

\begin{figure}[htbp]
  \centerline{\scalebox{0.70}{\rotatebox{\GnuplotRotatePS}{\includegraphics{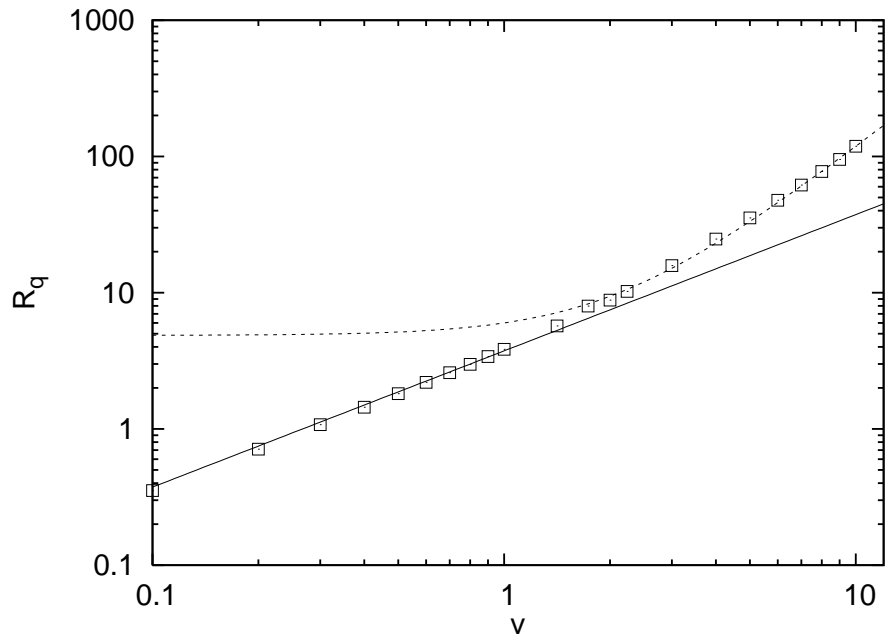}}}}
  \caption{\label{SokalRq}The radius $R_q$ for a two
    dimensional square lattice, i.e. $r=4$. The solid line is $f(v)
    \sim a\cdot v$ and the dashed line is $g(v) \sim a + b\cdot v^2$.}
\end{figure}

In a large paper by Alan Sokal\cite{Sokal:2001} it is shown that the
complex $q$ zeros of the partition function $Z(v,q)$ for $|1 + v| \le
1$ are all located within a circle given by the maximal degree of the
graph. The restriction $|1 + v| \le 1$ corresponds to the
\emph{antiferromagnetic} Potts model, which is not what we have
considered in this paper. If the restriction $|1 + v| \le 1$ is
relaxed the radius is found to scale as (for spatially constant $v$)
\begin{equation}
  \label{qRadius}
  R_q \sim \max \left[v , v^{r/2} \right], 
\end{equation}
where $r$ is the maximum degree of the graph, i.e. the maximum number
of edges incident on any one vertex. For an ordinary cubic lattice in
two dimensions we have $r=4$, hence we expect to see a crossover from
$v$ to $v^2$ scaling around $v = 1$. \Figref{SokalRq} shows the radius
$R_q$ as a function of $v$.

\section{Conclusion}
We have shown that the nontrivial information of the Potts model is
contained in the density $\Gamma_{\Graph{G}}({\Clusters},\Edges)$, and
this is independent of $q$. $\Gamma_{\Graph{G}}({\Clusters},\Edges)$
is \emph{purely} combinatorial/geometric property of the underlying
lattice, emphasizing the connection between these concepts and 
critical phenomena.  Furthermore we have devised an algorithm to
estimate $\Gamma_{\Graph{G}}({\Clusters},\Edges)$ from Monte Carlo
simulations, and used this to study various properties of the Potts /
Random Cluster model.

\end{document}